\documentclass[twocolumn,epsbox]{jpsj2}
\def\runtitle{Ferromagnetism and Superconductivity in the multi-orbital Hubbard Model}
\def\runauthor{Kazuhiro {\sc Sano}   and  Yoshiaki {\sc \=Ono}}

\newcommand{\simj}{\stackrel{>}{_\sim}}
\newcommand{\simk}{\stackrel{<}{_\sim}}

\title{%
Ferromagnetism and Superconductivity in the multi-orbital Hubbard Model: \\
Hund's Rule Coupling versus Crystal-Field Splitting
}

\author{%
Kazuhiro {\sc Sano}\thanks{E-mail address: sano@phen.mie-u.ac.jp} and  Yoshiaki {\sc \=Ono}\raisebox{0.5ex}{1}
}

\inst{%
Department of Physics Engineering,  Mie University, Tsu, Mie 514-8507       \\ 
\raisebox{0.5ex}{1}Department of Physics, Nagoya University, Nagoya 464-8602
}

\recdate{\today}

\abst{%
The multi-orbital Hubbard model in one dimension is studied using the numerical diagonalization method. Due to the effect of the crystal-field splitting $\Delta$, the fully polarized ferromagnetism which is observed in the strong coupling regime becomes unstable against the partially polarized ferromagnetism when the Hund's rule coupling $J$ is smaller than a certain critical value of order of $\Delta$. 
In the vicinity of the partially polarized ferromagnetism, the orbital fluctuation develops due to the competition between the Hund's rule coupling and the crystal-field splitting. The superconducting phase with the Luttinger liquid parameter $K_{\rho}>1$ is observed for the singlet ground state in this region. 
}

\kword{%
electronic structure, ferromagnetism, superconductivity, Luttinger liquid, Hubbard model
}

\begin{document}
\sloppy
\maketitle

%\section{Introduction}
Recently, the strongly correlated electron systems with orbital degrees of freedom have attracted much interest. Striking phenomena of such systems are the colossal magnetoresistance in manganites La$_{1-x}$Sr$_{x}$MnO$_3$ \cite{Tokura}, the triplet pairing superconductivity in the ruthenate Sr$_2$RuO$_4$ \cite{Maeno} and the metal-insulator transition in alkali-doped fullerides A$_x$C$_{60}$ \cite{Gunnarsson}. 

Another interesting class of materials is cobalt oxides such as La$_{1-x}$Sr$_{x}$CoO$_3$ \cite{Jonker,Heikes,Raccah} which shows the spin state transition. Because the Hund's rule coupling and the crystal-field splitting between the $t_{2g}$ and $e_g$ orbitals are close to each other, the spin state of the cobalt ion depends on temperature, doping concentration and crystal structure. For example, the ground state of the Co$^{3+}$ ($3d^6$) ion in LaCoO$_3$ is known to be a low-spin (LS) state ($t_{2g}^6 e_g^0$, $S=0$), while, with increasing temperature, the spin state gradually changes into an intermediate-spin (IS) state ($t_{2g}^5 e_g^1$, $S=1$) and/or a high-spin (HS) state ($t_{2g}^4 e_g^2$, $S=2$) \cite{Heikes,Raccah,Asai,Itoh1}. 
With Sr doping, La$_{1-x}$Sr$_{x}$CoO$_3$ shows a spin-glass for $x<0.18$ and a ferromagnetism for $x>0.18$ \cite{Louca}, where IS and/or HS states play crucial roles. 

The recent discovery of large thermoelectric power in Na$_{0.5}$CoO$_2$ \cite{Terasaki} has stimulated considerable attention on layered cobalt oxides such as Na$_{x}$CoO$_2$. Koshibae {\it et al.}  \cite{Koshibae} claimed that large degeneracy of electronic states due to a competition between the Hund's rule coupling and the crystal-field splitting, together with strong correlation plays a key role in Na$_{x}$CoO$_2$ as well as La$_{1-x}$Sr$_{x}$CoO$_3$. Weak ferromagnetism has been observed in Na$_{0.75}$CoO$_2$ \cite{Motohashi}. More recently, Takada {\it et al.} \cite{Takada} have discovered the superconductivity in Na$_{x}$CoO$_2\cdot y$H$_2$O with $T_c \approx 5K$ for $x \approx 0.35$ and $y \approx 1.3$. With the advent of the new findings, theoretical studies of the interplay of the Hund's rule coupling and the crystal-field splitting is highly desirable. 

The orbitally degenerate Hubbard model has been extensively investigated to clarify the effect of orbital degrees of freedom in the presence of the intra-atomic Coulomb interaction. Many authors \cite{Gill,Kuei,Held,Shen,Momoi,Sakamoto} have studied the ferromagnetism of this model and revealed that the Hund's rule coupling plays a crucial role in the ferromagnetism; however the effect of the crystal-field splitting was not considered there. Possible mechanisms of superconductivity have been proposed by several authors \cite{Takimoto,Imada,Yamaji,Nomura}, but the relationship with the ferromagnetism was not discussed there. 

In the present work, we investigate the ferromagnetism and the superconductivity in the multi-orbital Hubbard model, in particular paying attention to the effect of the interplay of the Hund's rule coupling $J$ and the crystal-field splitting $\Delta$. As the strong correlation effect plays crucial roles in ferromagnetism, a non-perturbative and reliable approach is required. We employ the numerical diagonalization method for the one-dimensional Hubbard model with finite system sizes. This approach has already been applied for the $\Delta=0$ case \cite{Gill,Kuei}. Although the available system size is fairly small, the results are in good agreement with the strong coupling analysis \cite{Kuei} and the results from the density-matrix renormalization-group method \cite{Sakamoto}. 
To examine the superconductivity, we calculate the critical exponent of the correlation functions $K_{\rho}$ based on the Luttinger liquid theory \cite{Haldane2,Voit}. The reliability of this approach has been extensively tested for various one-dimensional models such as the Hubbard model \cite{Schulz}, the $t$-$J$ model \cite{Ogata}, the $d$-$p$ model \cite{SanoPhysica1,SanoPRB,SanoPhysica2}, etc. We can thus expect that this approach is reliable for multi-orbital Hubbard model as well.

%\section{  Model and Luttinger Liquid Relation}
We consider the following Hamiltonian for the one-dimensional multi-orbital Hubbard  model:

\begin{eqnarray} 
H&=&-t\sum_{i,m,\sigma}(c_{i,m,\sigma}^{\dagger} c_{i+1,m,\sigma}+h.c.)  \nonumber \\
&+&U\sum_{i,m}n_{i,m,\uparrow}n_{i,m,\downarrow}
+U'\sum_{i,\sigma}n_{i,a,\sigma}n_{i,b,-\sigma} \nonumber \\
&+&(U'-J)\sum_{i,\sigma}n_{i,a,\sigma}n_{i,b,\sigma}
 +\frac{\Delta}{2}\sum_{i,\sigma}(n_{i,a,\sigma}-n_{i,b,\sigma})  \nonumber \\
 &-&J\sum_{i,m,\sigma} (c_{i,a,\uparrow}^{\dagger} c_{i,a,\downarrow}
 c_{i,b,\downarrow}^{\dagger} c_{i,b,\uparrow}+h.c.)\nonumber \\
 &-&J'\sum_{i,m,\sigma} (c_{i,a,\uparrow}^{\dagger} c_{i,a,\downarrow}^{\dagger} c_{i,b,\uparrow} c_{i,b,\downarrow}+h.c.)
   \end{eqnarray} 
where $c^{\dagger}_{i,m,\sigma}$ stands for creation operator of a electron with spin $\sigma$ in  the  orbital $m \ ( =a,b)$ at site $i$  and 
   $n_{i,m,\sigma}=c_{i,m,\sigma}^{\dagger}c_{i,m,\sigma}$.
Here, $t$ represents the hopping integral between the same orbitals and 
we set $t=1$ in this study.
The interaction parameters $U$, $U'$, $J$ and $J'$ stand  
 the intra- and inter-orbital direct Coulomb interactions, the exchange (Hund's rule) coupling and the pair-transfer, respectively.
 $\Delta$ stands the energy difference between the two atomic orbitals, i.e., the crystal-field splitting. 
For simplicity, we impose the relations, $J=J'$ and $U=U'+2J$, which holds exactly in $3d$-orbitals for $\Delta=0$ and is a good approximation for $\Delta \ne 0$. The model eq.(1) is schematically represented by Fig.1(a).

  To carry out a systematic calculation, we use the periodic boundary condition for $N_e=4m+2$ and the antiperiodic boundary condition for $N_e=4m$, where $N_e$ is the total electron number and $m$ is  an  integer. This choice of the boundary condition removes accidental degeneracy so that the ground state might always be a singlet with zero momentum except for the ferromagnetic state at large $J$. 
 The filling $n$ is defined  by  $n=N_{e}/N_{u}$, where $N_u$ is the total number of unit cells (each unit cell contains two orbitals).
  We numerically diagonalize the Hamiltonian eq.(1)  up to 14 sites (7 unit  cells)  using the standard Lanczos algorithm. 

     When the charge gap vanishes in the thermodynamic limit, the uniform  charge susceptibility $\chi_c$ is obtained from 
%\begin{equation}
$
\chi_c=\frac{4/N_u}{E_{0}(N_{e}+2,N_u)+E_{0}(N_{e}-2,N_u)-2E_{0}(N_{e},N_u)},
$ 
%\end{equation}
where $E_{0}(N_e,N_u)$ is the  ground state energy of a system with $N_u$ unit cells and $N_e$ electrons.
In the Luttinger liquid theory,  some relations  have been  established as universal relations in one-dimensional  models.\cite{Haldane2,Voit}  
 In  the model which is isotropic in spin space,  the critical exponents of various types of correlation functions are determined by a single parameter $K_\rho$.
  It is  predicted that the superconducting (SC) correlation  is dominant for $K_{\rho}>1$ (the correlation function decays   as $\sim r^{-(1+\frac{1}{K_{\rho}})}$), whereas  the CDW  or  SDW correlations are dominant  for $K_{\rho}<1$ (the correlation functions decay  as $\sim r^{-(1+K_{\rho})}$)
 in the Tomonaga-Luttinger regime.\cite{Voit}
%Here, the TL regime is characterized by a gapless spin excitation spectrum, while in the Luther-Emery regime, the spin excitation  has a gap.
The critical exponent $K_{\rho}$ is related to  the charge susceptibility $\chi_c$ and  the Drude weight $D$ by
$      K\sb{\rho}=\frac{1}{2}(\pi \chi_c D)^{1/2},$
with $D=\frac{\pi}{N_u} \frac{\partial^2 E_0(\phi)}{\partial \phi^2}$, where $E_0(\phi)$ is the total energy of the ground state as a function of  a magnetic flux $N_u \phi$.\cite{Voit}

\begin{figure}[t]
\begin{center}
\includegraphics[width=5cm]{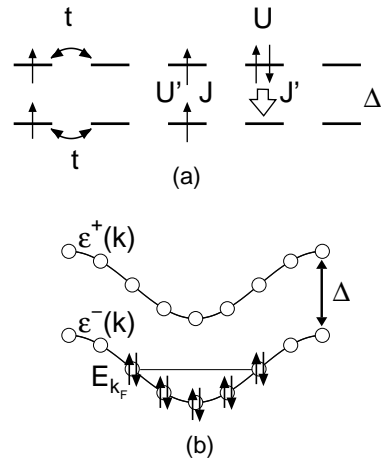}
\end{center}
\caption{Schematic diagrams of (a) the model Hamiltonian  and (b) the band structure in the noninteracting case.   
}
%\label{f1} 
\end{figure}

\begin{figure}[t]
\begin{center}
\includegraphics[width=7cm]{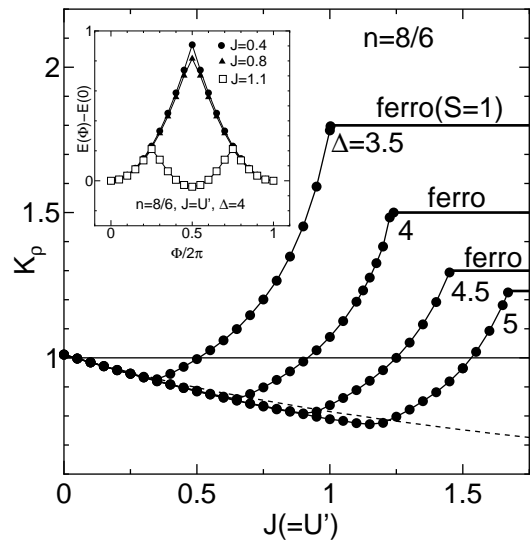}
\end{center}
\caption{ $K_{\rho}$ as a function of $J(=U')$ at $\Delta=3.5,4,4.5$and 5. The broken line represents a weak coupling estimation for $K_{\rho}$. Inset shows the energy difference $E_0(\phi)-E_0(0)$  as a function of an external flux $\phi$.}
%\label{f1}
\end{figure}

%\section{ Numerical Results}
  In the noninteracting case ($U=U'=J=0$), the Hamiltonian eq.(1) yields a dispersion relation  
$    \epsilon^{\pm}(k)=-2t \cos(k)\pm \frac{\Delta}{2} ,$
where $k$ is the wave vector and $\epsilon^+(k)$ $(\epsilon^-(k) )$ represents the upper (lower) band energy. 
When the lowest energy of the upper band, $\epsilon^+(0)$, is larger than the Fermi energy $E_{k_F}$,  electrons occupy only the lower band with 
 $k_F =\frac{\pi n}{2}$.
This  band structure  is schematically represented by Fig.1(b).
Hereafter, we mainly consider the case with  $\epsilon^+(0)>E_{k_F}$. 
%We  also assume that the  exponent $K_{\rho}$  is meaningful  still in the strong coupling regime.

Figure 2 shows  the value of $K_{\rho}$ as a function of $J(=U')$  for several values of $\Delta$ at the electron density  $n=8/6$.
As $J$ increases $K_{\rho}$  decreases for small $J$, while it increases for large $J$ and, then,  becomes larger than unity. 
In the region $K_{\rho}>1$, the SC correlation is expected to be most dominant compared to the CDW and SDW correlations. 
%Figure 2 also indicates that the enhancement of $K_{\rho}$ decreases as $\Delta$ increases. 
When $J$ is larger than a certain critical value,  the ground state changes into the partially ferromagnetic state with S=1 from the singlet state. 
%Roughly speaking, the transition into ferromagnetism  is caused by  the  Hund coupling $J$ in cooperation with $U'$ and $U$.

In Fig.2 the broken line represents a weak coupling estimation for $K_{\rho}$, where $E_0$ is calculated within the first order perturbation (Hartree approximation) \cite{Sano1,Sano2}. This  approximation shows a good agreement with the numerical result in the weak coupling regime, ensuring the small finite-size effect of the numerical calculation. In this approximation, the ground state energy is obtained  by the expectation value of the Hamiltonian eq.(1) using the noninteracting ground state where the lower band is exclusively occupied as shown in Fig.1(b). 
Then the effect of the upper band is omitted and the result is independent of $\Delta$ as shown in Fig.2. 
This approximation breaks down when the effect of the upper band becomes crucial. In this regime, $K_{\rho}$ rapidly increases with increasing $J$, and finally becomes larger than unity showing the superconducting state.

The critical values  of $J$ with $K_{\rho}=1$  are  $J_c \approx$ 0.5, 0.9, 1.3 and 1.6 for $\Delta=$ 3.5, 4, 4.5 and 5, respectively.  We  find that the phenomenological equation $J_c=\epsilon^+(0)-E_{k_F}$ can approximately lead to the above results.  Here, $\epsilon^+(0)-E_{k_F}$ corresponds to the lowest energy of the single-particle excitation from the lower band to the upper band. This equation is found to be a good approximation for various filling $n$. 
When $J$ exceeds $\epsilon^+(0)-E_{k_F}$, the inter-band excitation develops and the occupation number of the upper band increases, which results in the large orbital fluctuation accompanied by the fluctuation between the low-spin and the high-spin states. 
The mechanism of the superconductivity is related to this orbital fluctuation. 

We note that the superconductivity is also observed for the $J'=0$ ($J\ne 0$) case in contrast to the previous study \cite{Imada,Yamaji} where the pair-transfer $J'$ is crucial for the superconductivity. 
The importance of the upper band has also been pointed out in different types of multi-band models such as the $d$-$p$ model \cite{SanoPhysica1,SanoPRB,SanoPhysica2,Ono1} and the Hubbard ladder model \cite{Sano1,Sano2}. 

To confirm the superconducting state, we calculate the lowest energy of the singlet state $E_0(\phi)$   as a function of an external flux $\phi$.
As shown in the inset of Fig.2, the anomalous flux quantization occurs at $J=1.1$, where $K_{\rho}$ is about 1.4. When $J=0.8$ and $0.4$ , $K_{\rho}$ is less than  unity and  the anomalous flux quantization is not found. 
%The result also indicates that the superconductivity appears for large  $J$ at $K_{\rho}> 1$.

\begin{figure}[t]
\begin{center}
\includegraphics[width=6.6cm]{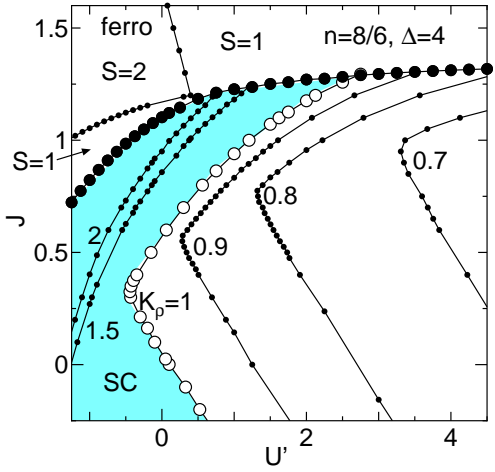}
\end{center}
\caption{Phase diagram of the superconducting state with $K_{\rho}>1$ (shadowed region) and the ferromagnetic state with $S\ne 0$ on the  $U'-J$ parameter plane with contour map of $K_\rho$.}
%\label{f1}
\end{figure}
\begin{figure}[t]
\begin{center}
\includegraphics[width=6.8cm]{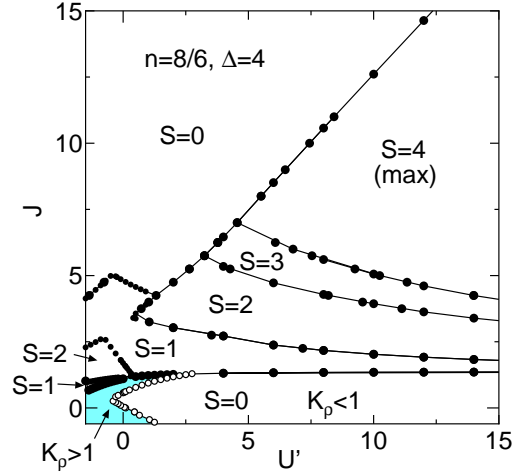}
\end{center}
\caption{Global phase diagram on the  $U'-J$ plane.}
%\label{f1}
\end{figure}

In Fig.3, we show the  phase diagram of the superconducting state with $K_{\rho}>1$  together with the ferromagnetic state with $S\ne 0$  on the $U'$ vs. $J$ parameter plane, where the value of $K_{\rho}$ is given by the contour map. 
The superconducting phase appears near the partially polarized  ferromagnetic  region. It extends from the attractive  region with $J<0$ and $U'<0$ to  the realistic parameter region for $3d$ transition-metals with $U'>J>0$. 
We have confirmed that the superconducting region increases as $\Delta$ decreases as shown in Fig.2.

Fig.4 shows the global phase diagram on the  $U'-J$ plane. When  $U'\simj J \simj \Delta$, the fully polarized ferromagnetism with $S=S_{\rm max}$  appears. It  accompanies the partially polarized ferromagnetism with $0<S<S_{\rm max}$ for $J\simk \Delta$. 
In the $\Delta=0$ case, the  ferromagnetism of the degenerate Hubbard model has been previously studied \cite{Gill,Kuei,Held,Shen,Momoi,Sakamoto}. 
In  one dimension, some rigorous results are shown in  the strong coupling limit $U\to \infty$: the ground state  is fully polarized ferromagnetism for $0<n<2$  except for $n=1$ when  $U'$ and $J=J'$ are positive and finite \cite{Shen}.  
Numerical result  suggests that the ferromagnetism is stable also for $n=1$ in the  strong coupling region \cite{Sakamoto}.  
In infinite dimensions, the dynamical mean-field theory shows the existence of the ferromagnetism in the same parameter region observed in one dimension \cite{Momoi}. 
The ferromagnetism for $n\ne 1$ is found to be metallic and mainly caused by the double-exchange mechanism \cite{Sakamoto}, which is also the present case with $\Delta\ne 0$. 
The existence of the partially polarized ferromagnetism for $\Delta>0$ has been reported in a different type of the two-band Hubbard model \cite{Kusakabe}. 
The ferromagnetic phase for $U' \simk 0$ is complicated as shown in Fig.4. The origin of this phase is not clear at this stage. 

\begin{figure}[t]
\begin{center}
\includegraphics[width=6.4cm]{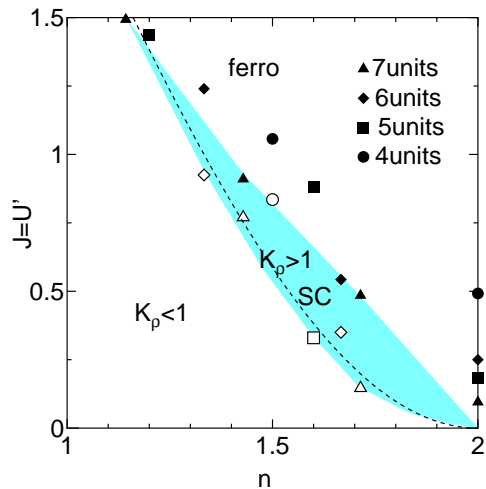}
\end{center}
\caption{Phase diagram on the $n-J$ plane. The broken line represents 
 the phenomenological equation $J_c=\epsilon^+(0)-E_{k_F}$ (see text) as a function of $n$.}
\label{f1}
\end{figure}

 In Fig.5, we show  the doping dependence of the critical values of $J$ for the superconductivity and the ferromagnetism  at  $\Delta=4$ with $J=U'$, 
where we use $n$=8/6, 10/7, 6/4, 8/5 and 12/7 systems. 
The critical values  are defined above which $K_{\rho}>1$ for the superconductivity and $S>0$ for the ferromagnetism, respectively. 
Although the finite size effect is considerably large,  the phase  boundary for the superconductivity   can be approximately given by  the   phenomenological equation, $J_c=\epsilon^+(0)-E_{k_F}$, mentioned before. 
 
For $n=8/7$ and $n=6/5$, we cannot use the expression 
$K\sb{\rho}=\frac{1}{2}(\pi \chi_c D)^{1/2}$, 
because the system  has a charge gap at $N_e=N_u$ and, then, $\chi_c$ is not obtained from the finite size systems. 
Instead we use another expression, 
$K\sb{\rho}=\frac{D}{2v_c}$ with the charge velocity $v_c$. 
Although there is a larger finite size effect, we find $K_{\rho}<1$ until the ferromagnetic transition occurs. 
Furthermore, the result of  the  flux quantization also shows no sign of the superconductivity in this case. 
These results suggest that the superconducting phase may disappear for $ n \simk 6/5$.

%\section{Summary and Discussion}
In conclusion, we have obtained the phase diagram of the one-dimensional Hubbard model with two-fold orbital degeneracy.
The fully polarized ferromagnetism has been found in the strong coupling regime with $U'\simj J\simj \Delta$. 
For $1<n<2$, the ferromagnetism is metallic and mainly caused by the double-exchange mechanism \cite{Sakamoto}. 
The crystal-field splitting destroys the fully polarized ferromagnetism resulting in the partially polarized one for $J\simk \Delta$. 
In the vicinity of the partially polarized ferromagnetism, we have found the superconducting phase, when $J$ exceeds the lowest energy of the inter-band excitation, which extends to the realistic parameter region for $3d$ transition-metals with $U'>J$.

Sakamoto {\it et al.} \cite{Sakamoto} claimed that the metallic ferromagnetism  appears in a similar parameter region in any dimension  by comparing the results from one dimension with those from infinite dimensions.
It is natural to think  that this ferromagnetism  will appear in two and three dimensions even in the presence of $\Delta$. 
Then, we expect that a partially polarized (weak) ferromagnetism appears in  real materials, in which the Hund's rule coupling and the crystal-field splitting compete to each other, such as cobalt oxides. 
In fact a weak ferromagnetism has been observed in the layered Na$_{0.75}$CoO$_2$ \cite{Motohashi} as well as the perovskite R$_{1-x}$A$_{x}$CoO$_3$ \cite{Sato}.

Finally we make some comments on the superconductivity in Na$_{x}$CoO$_2\cdot y$H$_2$O \cite{Takada}. Although the several authors \cite{Baskaran,Shastry,Lee,Ogata2} have discussed the system using the non-degenerate $t$-$J$ model, the orbital degeneracy of 3$d$ electrons is considered to play a crucial role in Na$_{x}$CoO$_2$ as well as La$_{1-x}$Sr$_{x}$CoO$_3$ \cite{Koshibae}. The competition between the Hund's rule coupling and the crystal-field splitting causes the large orbital fluctuation, accompanied by the fluctuation between the low-spin and the high-spin states at each Co ion, which mediates the superconductivity. As the orbital fluctuation has a local character, the mechanism for the superconductivity could be common in all dimensions. We hence expect that the multi-orbital mechanism is responsible for the superconductivity in layered Na$_{x}$CoO$_2\cdot y$H$_2$O. Exploration of the superconductivity in the vicinity of the weak ferromagnetism in the perovskite R$_{1-x}$A$_{x}$CoO$_3$ \cite{Sato} may be promising.

We would like to thank Prof. Masatoshi Sato who led our attention to cobalt oxides and suggested the importance of the Hund's rule coupling on the superconductivity. 
This work was partially supported by the Grant-in-Aid for Scientific Research from the Ministry of Education, Science, Sports and Culture.


\begin{thebibliography}{99}

\bibitem{Tokura} For a review, see, for example, Y. Tokura, Ed.: {\it Colossal Magnetoresistive Oxides} (Gordon and Breach Science, New York, 2000).
%M. B. Salamon and M. Jaime, Rev. Mod. Phys. {\bf 73} (2001) 583.
%M. Imada, A. Fujimori, and Y. Tokura: Rev. Mod. Phys. {\bf 70} (1998) 1039.

\bibitem{Maeno} Y. Maeno, H. Hashimoto, K. Yoshida, S Nishizaki, T. Fujita, J. G. Bednorz and F. Lichtenberg: Nature (London) {\bf 372} (1994) 532.

\bibitem{Gunnarsson} O. Gunnarsson, E. Koch and R. M. Martin: Phys. Rev. B {\bf 54} (1996) R11026.
%O. Gunnarsson: Rev. Mod. Phys. {\bf 69} (1997) 575.

%\bibitem{Koch} E. Koch, O. Gunnarsson and R. M. Martin: Phys. Rev. B {\bf 60} (1999) 15714.

\bibitem{Jonker} G. H. Jonker and J. H. Van Santen: Physica {\bf 19} (1953) 120.
\bibitem{Heikes} R. R. Heikes, R. C. Miller and R. Mazelsky: Physica {\bf 30} (1964) 1600.

\bibitem{Raccah} P. M. Raccah and J. B. Goodenough: Phys. Rev. {\bf 155} (1967) 932.

\bibitem{Asai} K. Asai, P. Gehring, H. Chou, and G. Shirane: Phys. Rev. B {\bf 40} (1989) 10982.

\bibitem{Itoh1} M. Itoh, M. Sugahara, I. Natori and K. Motoya: J. Phys. Soc. Jan. {\bf 64} (1995) 3967.

%\bibitem{Itoh2} M. Itoh, I. Natori, S. Kubota and K. Motoya: J. Phys. Soc. Jpn. {\bf 63} (1994) 1486.

\bibitem{Louca} D. Louca, J. L. Sarrao, J. D. Thompson, H. Roder and G. H. Kwei: Phys. Rev. B {\bf 60} (1999) 10378.


\bibitem{Terasaki} I. Terasaki, Y. Sasago and K. Uchinokura, Phys. Rev. B {\bf 56} (1997) R12685.

\bibitem{Koshibae} W. Koshibae, K. Tsutsui and S. Maekawa, Phys. Rev. B {\bf 62} (2000) 6869.

\bibitem{Motohashi} T. Motohashi, R. Ueda, E. Naujalis, T. Tojo, I. Terasaki, T. Atake, M. Karppinen and H. Yamauchi: Phys. Rev. B {\bf 67} (2003) 064406. 

\bibitem{Takada} K. Takada, H. Sakurai, E. Takayama-Muromachi, F. Izumi, R. A. Dilanian and T. Sasaki: Nature {\bf 422} (2003) 53. 

%\bibitem{Bunemann} J. B\"unemann, W. Weber and F. Gebhard: Phys. Rev. B {\bf 57} (1998) 6896.

%\bibitem{Motome} Y. Motome and M. Imada: J. Phys. Soc. Jan. {\bf 67} (1998) 3199.

%\bibitem{Han} J. E. Han, M. Jarrell and D. L. Cox: Phys. Rev. B {\bf 58} (1998) R4199.

%\bibitem{Koga} A. Koga, Y. Imai, and N. Kawakami: Phys. Rev. B {\bf 66} (2002) 165107.

%\bibitem{Ono} Y. \=Ono, M. Potthoff and R. Bulla: Phys. Rev. {\bf B 67} (2003) 035119.

%\bibitem{Okabe} T. Okabe: Prog. Theor. Phys. {\bf 98} (1997) 33.

\bibitem{Gill} W. Gill and D. J. Scalapino: Phys. Rev. B {\bf 35} (1987) 215.

\bibitem{Kuei} J. Kuei and R. T. Scalettar: Phys. Rev. B {\bf 55} (1997) 14968.

%\bibitem{Hirsch} J. E. Hirsch: Phys. Rev. B {\bf 56} (1997) 11022.

\bibitem{Held} K. Held and D. Vollhardt: Eur. Phys. J. B {\bf 5} (1998) 473.

\bibitem{Shen} S. Q. Shen: Phys. Rev. {\bf B57}  (1998) 6474.

\bibitem{Momoi} T. Momoi and K. Kubo: Phys. Rev. B {\bf 58} (1998) R567.

\bibitem{Sakamoto}  H. Sakamoto, T. Momoi and K. Kubo: Phys. Rev. B {\bf 65} (2002) 224403.

\bibitem{Takimoto}  T. Takimoto: Phys. Rev. B {\bf 62} (2000) R14641.

\bibitem{Imada} M. Imada: J. Phys. Soc. Jpn. {\bf 70} (2001) 1218.

\bibitem{Yamaji} K. Yamaji: J. Phys. Soc. Jpn. {\bf 70} (2001) 1476.

\bibitem{Nomura} T. Nomura and K. Yamada: J. Phys. Soc. Jan. {\bf 71} (2002) 1993.

%\bibitem{Haldane} F.D.M. Haldane:  Phys. Rev. Lett. {\bf 45} (1980) 1358.

\bibitem{Haldane2} F.D.M. Haldane: J. Phys. {\bf C14}  (1981) 2585.

\bibitem{Voit} J. Voit: Rep. Prog. Phys. {\bf 58}  (1995) 977.

\bibitem{Schulz} H. J. Schulz: Phys. Rev. Lett. {\bf 64} (1990) 2831. 

\bibitem{Ogata} M. Ogata, M. U. Luchini, S. Sorella and F. F. Assaad: Phys. Rev. Lett. {\bf 66} (1991) 2388. 

\bibitem{SanoPhysica1} K. Sano and Y. \={O}no: Physica {\bf C205} (1993) 170.
 
\bibitem{SanoPRB} K. Sano and Y. \={O}no: Phys. Rev. {\bf B51}  (1995) 1175.

\bibitem{SanoPhysica2} K. Sano and Y. \={O}no: Physica {\bf C242}  (1995) 113.

\bibitem{Sano1} K. Sano: Physica {\bf B281\&282}  (2000) 829.

\bibitem{Sano2} K. Sano: J. Phys. Soc. Jpn. {\bf 69} (2000) 1000.

\bibitem{Ono1} Y. \=Ono and K. Sano: J. Phys. Soc. Jpn. Suppl. {\bf 71} (2002) 356. 


\bibitem{Kusakabe} K. Kusakabe, S. Watanabe and Y. Kuramoto: J. Phys. Soc. Jpn. Suppl. {\bf 71} (2002) 311.

%\bibitem{chi1}
%The $\chi_c$ defined by eq.(2) is meaningless in this case  because the system  has a charge gap at  the quarter-filling($N_e=N_u$). 

%\bibitem{chi2}
%  The value of $K_{\rho}$ is  also calculated by the equation 
%$    K\sb{\rho}=\frac{D}{2v_c}$, where $v_c$ is the charge velocity.
%It does not need $\chi_c$, however, the finite size effect of the above equation is larger than that of eq. (3).

%\bibitem{Zener} C. Zener: Phys. Rev. {\bf 81} (1951) 440. 

\bibitem{Baskaran}  G. Baskaran: cond-mat/0303649.

\bibitem{Shastry} Brijesh Kumar and B. Sriram Shastry: cond-mat/0304210.

\bibitem{Lee} Q.-H. Wang, D.-H. Lee and P. A. Lee: cond-mat/0304377.

\bibitem{Ogata2} M. Ogata: cond-mat/0304405.

\bibitem{Sato} H. Masuda, T. Fujita, T. Miyashita, M. Soda, Y. Yasui, Y. Kobayashi and M. Sato: J. Phys. Soc. Jpn. {\bf 72} (2003) 873. 



\end{thebibliography}
\end{document}